\documentclass[fleqn,usenatbib]{mnras}

\usepackage{newtxtext,newtxmath}

\usepackage[T1]{fontenc}

\DeclareRobustCommand{\VAN}[3]{#2}
\let\VANthebibliography\thebibliography
\def\thebibliography{\DeclareRobustCommand{\VAN}[3]{##3}\VANthebibliography}

\usepackage{graphicx}	
\usepackage{amsmath}	

\usepackage{longtable}

\title[Youth and IR Excess of SIPS J2045-6332,]{Unveiling the Infrared Excess of SIPS J2045-6332: Evidence for a Young Stellar Object with Potential Low-Mass Companion}

\author[]{\href{https://orcid.org/0000-0001-8343-0820}{Michiharu Hyogo}$^{1,2}$\thanks{E-mail:michiharu.hyogo77@gmail.com},
    \href{https://orcid.org/0000-0003-2235-761X}{Thomas P. Bickle}$^{3}$,
    Joseph R. Biggs$^{1}$,
    \href{https://orcid.org/0000-0002-6523-9536}{Adam J. Burgasser}$^{4}$,
    \href{https://orcid.org/0000-0001-7896-5791}{Dan Caselden}$^{5}$,
    \newauthor
    \href{https://orcid.org/0000-0003-2478-0120}{Sarah Casewell}$^{6}$,
    Sergio B. Dieterich$^{6}$,
    \href{https://orcid.org/0000-0002-4143-2550}{Hugo A. Durantini Luca}$^{1,2,7}$,
    \href{https://orcid.org/0000-0001-6251-0573}{Jacqueline Faherty}$^{5}$,
    \newauthor
    \href{https://orcid.org/0000-0001-8170-7072}{Daniella Bardalez Gagliuffi}$^{8}$,
    \href{https://orcid.org/0000-0002-2592-9612}{Jonathan Gagné}$^{9,10}$,
    \href{https://orcid.org/0000-0003-4269-260X}{J.\ Davy Kirkpatrick}$^{11}$,
    \href{https://orcid.org/0000-0002-2387-5489}{Marc J. Kuchner}$^{12}$,
    {Carey M. Lisse}$^{13}$,
    \newauthor
    \href{https://orcid.org/0000-0001-7519-1700}{Federico Marocco}$^{11}$,
    \href{https://orcid.org/0000-0002-1125-7384}{Aaron M. Meisner}$^{14}$,
    \href{https://orcid.org/0000-0003-4083-9962}{Austin Rothermich}$^{5,15}$,
    \href{https://orcid.org/0000-0002-6294-5937}{Adam C. Schneider}$^{16}$,
    Steven M. Silverberg$^{17}$,
    \newauthor
    Disk Detective Collaboration, 
    AND The Backyard Worlds: Planet 9 Collaboration
\\
$^{1}$ Disk Detective Collaboration, USA \\
$^{2}$ Backyard Worlds: Planet 9, USA \\
$^{3}$ School of Physical Sciences, The Open University, Milton Keynes, MK7 6AA, UK\\
$^{4}$ Department of Astronomy \& Astrophysics, UC San Diego, La Jolla, CA, USA\\
$^{5}$ Department of Astrophysics, American Museum of Natural History, Central Park West at 79th Street, NY 10024, USA\\
$^{6}$ Centre for Exoplanet Research, School of Physics and Astronomy University of Leicester, University Road, Leicester, LE1 7RH, UK\\
$^{7}$ UNC-Famaf, Av. Medina Allende s/n, Ciudad Universitaria, X5000 HUA, Cordoba, Argentina\\
$^{8}$ Department of Physics \& Astronomy, Amherst College, 25 East Drive, Amherst, MA 01003, USA\\
$^{9}$ Planétarium Rio Tinto Alcan, 4801 Pierre-de Coubertin Ave, Montreal, Quebec H1V 3V4, Canada\\
$^{10}$ Institute for Research on Exoplanets, Université de Montréal, Département de Physique, C.P. 6128 Succ. Centre-ville, Montréal, QC
H3C 3J7, Canada\\
$^{11}$ IPAC, Mail Code 100-22, California Institute of Technology, 1200 E. California Blvd., Pasadena, CA 91125, USA\\
$^{12}$ NASA Goddard Space Flight Center, Exoplanets and Stellar Astrophysics Laboratory, Code 667, Greenbelt, MD 20771, USA\\
$^{13}$ Johns Hopkins University Applied Physics Laboratory  located at 11100 Johns Hopkins Road, in Laurel, MD 20723 USA\\
$^{14}$ NSF's National Optical-Infrared Astronomy Research Laboratory, 950 N. Cherry Ave., Tucson, AZ 85719, USA\\
$^{15}$ Department of Physics, Graduate Center, City University of New York, 365 5th Ave., New York, NY 10016, USA\\
$^{16}$ United States Naval Observatory, Flagstaff Station, 10391 West Naval Observatory Road, Flagstaff, AZ 86005, USA\\
$^{17}$ Smithsonian Astrophysical Observatory, 60 Garden St., Cambridge, MA 01238, USA\\
}

\date{Accepted XXX. Received YYY; in original form ZZZ}

\pubyear{\the\year{}}

\begin{document}
\label{firstpage}
\pagerange{\pageref{firstpage}--\pageref{lastpage}}
\maketitle

\begin{abstract}
The Disk Detective project, a citizen science initiative, aims to identify circumstellar discs around stars by detecting objects with infrared (IR) excess using data from the Wide-field Infrared Survey Explorer (WISE). In this study, we investigate SIPS J2045-6332, a potential brown dwarf with significant IR excess in WISE and 2MASS bands, initially identified by project volunteers. Despite early indicators of a circumstellar disc, discrepancies between observed brightness and expected Spectral Energy Distribution (SED) models suggested unusual properties. To explore potential explanations, we created SED templates for spectral types M9 to L4 and compared them with SIPS J2045-6332’s photometric data, revealing an excess brightness that points to either an unresolved low-mass companion or a young, inflated primary star. Further analysis of infrared spectral features and surface gravity indicators supports a youthful classification, estimating the object's age at 26–200 million years. Observations also suggest the presence of a mid L-type companion at a projected distance of 6.7 AU. This study highlights SIPS J2045-6332 as an intriguing system with unique IR characteristics and recommends follow-up observations with high-resolution telescopes to confirm the companion hypothesis and further characterise the system.
\end{abstract}




\section{Introduction}

The Disk Detective project, a citizen science initiative, is dedicated to identifying circumstellar discs by analysing all-sky Wide-field Infrared Survey Explorer (WISE) images (see \citealt{kuchner2016}; \citealt{silverberg2017,silverberg2018,silverberg2020}; \citealt{schutte2020}; \citealt{higashio2022}; \citealt{laos2022}). With the active participation of volunteers, the project aims to locate objects that exhibit significant infrared excess, a key indicator of circumstellar discs.  A significant challenge in identifying Young Stellar Object (YSO) discs and debris discs is the frequent occurrence of confusion and contamination, which limits the effectiveness of searches using the Wide-field Infrared Survey Explorer (WISE) (\citealt{wright2010}). The point spread function (PSF) in the W3 band has a full width at half maximum (FWHM) of 7.5 arcseconds, resulting in a high probability of multiple point sources—such as background stars, galaxies, and nebulae—within the PSF. Additionally, image artefacts, including diffraction spikes, latent images, and optical ghosts, can further contaminate the field of view within the same PSF.

To address these issues, we analysed imaging data from the Pan-STARRS Data Archive \citep{chambers2016}, the SkyMapper Southern Sky Survey \citep{onken2024}, and the Two Micron All Sky Survey (2MASS) \citep{cutri2003}. Volunteers from the citizen science project Disk Detective used these images, alongside those from the Wide-field Infrared Survey Explorer (WISE), to identify potential background sources through the Disk Detective website (http://www.diskdetective.org). Using these datasets, they checked for additional sources within a radius of 7.5 arcseconds.

Subsequently, the AllWISE Data Release \citep{cutri2021} was accessed to retrieve available W1 and W3 magnitudes. Objects with W1–W3 > 0.25 were flagged as candidates with infrared (IR) excess, suggestive of potential circumstellar discs. Once these criteria were met, volunteers submitted the flagged objects for further investigation to confirm disc characteristics. From the initial Disk Detective version 2.1 dataset, this process yielded 1,419 candidate objects indicative of potential circumstellar discs.

Following this process, selected volunteers conducted additional analyses to confirm the presence of discs around the submitted candidates. The primary parameter assessed was the infrared (IR) excess criterion, as defined by \cite{avenhaus2012}. In this study, IR excess is quantified by the following equation:

\begin{equation}
{
\sigma'_i = \frac{e_i}{\sigma_i} \frac{1}{\sqrt{\chi^2}}
}
\end{equation}

where $e_i$ represents the difference between the observed WISE1 - WISE3 colour and the intrinsic colour of the objects, $\sigma_i$ denotes the uncertainty in the WISE1 - WISE3 colour, and $\chi^2$ is the goodness-of-fit metric, available within table A.1 of the same study. The intrinsic colours of the objects were empirically determined by fitting a fourth-order polynomial, expressed as a function of absolute WISE1 magnitude, using a sample of nearby M-type stars from the RECONS programme (\citealt{jao2005}; \citealt{henry2006}), along with data from the 2MASS and WISE surveys. Upon assessing this IR excess criterion, a comprehensive literature review was conducted to confirm that no prior studies had reported discs associated with these objects.

In this analysis, we identified SIPS J2045-6332 as exhibiting an unusually high IR excess, with a significance level of $\sigma'_i > 15$, the largest among the 1,419 candidates surveyed. Notably, this object has the fourth-latest spectral type in the sample, suggesting it may be a promising disc candidate surrounding a brown dwarf. However, as discussed in the following sections, this object is unusually bright not only in the WISE3 band but also in the 2MASS J, H, and K bands compared to the SED templates of the same spectral type—a rare characteristic for a brown dwarf with circumstellar discs. Several studies suggest plausible explanations for this anomaly, one of which is the possibility of a low-mass mid-L-type companion in the system \citep{deacon2017}.

Previous studies have identified a low-mass companion associated with the Vega-like system HD 155826, characterised by significant infrared (IR) excess in the wavelength ranges of 12–25 µm and 12–60 µm (e.g. \citealt{lisse2002}). This low-mass companion was located at an angular separation of approximately 7" from HD 155826 and was discovered through follow-up photometric imaging conducted using the JPL MIRLIN camera. Based on these observations, the study concluded that a circumstellar disc is not required to explain the observed IR excess. There exists the possibility that SIPS J2045-6332 shares similar physical characteristics. Alternatively, as suggested by \citet{galvez-ortiz2014} and \citet{marocco2013}, the system may be a young, low-mass object. The primary objective of the present study is to further investigate the physical properties identified in these previous works.

In the second section, we provide an overview of the fundamental physical properties of the system under study, including its distance, celestial coordinates, estimated age, and other notable characteristics, such as the potential existence of a low-mass companion. The third section details the methodology used to construct spectral energy distribution (SED) templates for spectral types ranging from M9 to L4. In the fourth section, we conduct a comparative analysis between these constructed SED templates and the observed photometric data for SIPS J2045-6332, aiming to elucidate the reasons behind its unusually high photometric brightness relative to SED predictions.

\section{Physical properties and Discoveries of SIPS J2045-6332}

The celestial coordinates of the object are RA = 311.26028667400 and DEC = -63.53606845258. Its proper motion is measured at 78.7 mas/yr in the RA direction and -216.0 mas/yr in the DEC direction, with a parallax of 44.2 mas \citep{gaia2021}. The object is classified as spectral type M9.0 $\pm$ 0.5 \citep{reid2008}. Based on its lithium abundance, estimated at $\log N(\text{Li}) = 3.5 \pm 0.5$ dex \citep{galvez-ortiz2014}, the object's age is inferred to be between 7 and 100 million years. This information is summarised in Table 1.

Beyond these data, several studies have suggested the potential presence of a low-mass companion. For instance, \citet{deacon2017} used the image shape measurement technique developed by \citet{hoekstra2005} to identify visual binary stars, revealing that SIPS J2045$-$6332 has a total ellipticity exceeding 0.5. Additionally, \citet{marocco2013} detected spectral anomalies within this system, which could indicate the presence of an unresolved companion. This physical characteristic of the system will be further investigated in this study.

\subsection{Infrared Spectroscopic Observations and Data Acquisition}
To further assess the age of this system, infrared (IR) spectra were obtained using the FIRE Echelle instrument on the 6.5 m Magellan telescopes. Observations were carried out on 3 August 2015 (UT), covering a wavelength range from 0.82 µm to 2.51 µm with a HAWAII-2RG infrared focal plane array. The airmass was 1.215, and the spectral resolution achieved was R = 6000, using a 0.60" slit. The IR spectra are presented in Figure 1.

The spectra were reduced using a custom version of the FIREHOSE pipeline, based on the MASE pipeline (\citealt{bochanski2009}), written in the Interactive Data Language (IDL). The SpexTool (\citealt{vacca2003}) was used to eliminate bad pixels and detector hot spots. The data were also flat-fielded and then extracted using a FIREHOSE algorithm.

\begin{table*}
\centering
\normalsize 
\caption{Physical properties of the objects considered in this study}
\label{Table1}
\hspace*{-1cm} 
\setlength{\tabcolsep}{15pt} 
\renewcommand{\arraystretch}{1.2} 
\resizebox{1.05\textwidth}{!}{
\begin{tabular}{|l|c|c|c|c|c|c|}
\hline
\textbf{Name} & \textbf{Absolute G mag} & \textbf{RA, Dec (degrees)} & \textbf{Plx (mas)} & \textbf{Age (Myr)} & \textbf{Proper Motion (mas/yr)} & \textbf{Banyan Probability} \\
\hline
SIPS J2045$-$6332 & 15.872 $\pm$ 0.009 & 311.26076945279, -63.53611701456 & 44.2$\pm$0.2 & $7--100^{1}$ & 78.7$\pm$0.1, -216.0$\pm$0.2 & 99.9\% $Field Star^{2}$ \\
Gaia DR3 6450595924275742848 & 17.91 $\pm$ 0.03& 311.26028667400, -63.53606845258 & 44.9$\pm$0.6 & $7--100^{1}$ & 78.1$\pm$0.4, -223 $\pm$1 & 99.9\% $Field Star^{2}$ \\
\hline
\end{tabular}%
}
\raggedright
\footnotesize \textbf{References.} All data, with the exception of age and Banyan Probability, were sourced from \cite{gaia2021}. 1. Derived from \cite{galvez-ortiz2014}. 2. \cite{gange2018}
\end{table*}

\section{SED MODEL}

To characterise the near-infrared excess associated with the object SIPS J2045-6332, we constructed spectral energy distribution (SED) templates. Data were compiled from single, field-age dwarfs using observations obtained from instruments such as SpeX on the Infrared Telescope Facility (IRTF; \citealt{rayner2008}, \citealt{cushing2005}), NIRSPEC on the Keck Observatory (\citealt{mclean2003,mclean2007}), and the Infrared Spectrograph (IRS) on the Spitzer Space Telescope (\citealt{houck2004}).

The near-infrared spectra from the SpeX and NIRSPEC spectrometers span a wavelength range of approximately 0.8 to 2.0 µm, while the Spitzer IRS provides coverage from approximately 5.0 to 15 µm. Spectral data beyond 15 µm (extending up to 35 µm) were excluded from the analysis, as not all objects in the sample were observed across this range.

As an illustration, in constructing templates for the M9 spectral type, we utilised astrophysical objects denoted as LHS 2065, 2MASS J23515044-2537367, 2MASS J00242463-0158201, and SSSPM J1013-1356. The pertinent physical characteristics of these objects are detailed in Table 2. Additionally, Table 2 provides information on the physical characteristics of objects corresponding to other spectral types, spanning from L0 to L4.

In constructing spectral energy distribution (SED) templates, a systematic methodology was implemented, comprising the following steps:

\begin{itemize}

\item The initial step involved applying filter corrections to the spectra of individual low-mass, late-type M stars and L-type stars. To achieve this, we developed a Python script that reads the spectrum of each object and loads filter profiles from various surveys, including the 2MASS H, J, and K bands, obtained from the SVO online repository (\citealt{rodrigo2012}, \citealt{rodrigo2020}). The script then convolves the object's spectrum with the corresponding filter profile, and the magnitude of the corrected flux is computed. Next, we calculated the ratio of the computed flux to the flux values documented in the 2MASS and ALLWISE catalogs, accessible through VizieR. Specifically, we computed the fluxes in the 2MASS J, H, and K bands, and, using the corresponding photometric values from the 2MASS catalog, determined the ratios between the observed and synthetic fluxes in these bands. The average of these ratios was then computed and used as a multiplicative factor, which was systematically applied to the original spectra for the required adjustments.

\item The next phase involved the computation of absolute magnitudes corresponding to specific wavelengths. Parallax values for each object were sourced from the GAIA eDR3 catalog, facilitating the determination of absolute magnitudes across each spectral band. This consistent methodology was applied to all gathered objects, contributing to the construction of the spectral energy distribution (SED) templates.

\item The final step involves determining the median brightness for each wavelength across all observed objects. This procedural measure is implemented to effectively mitigate the impact of outliers. Despite meticulous efforts to include only aged, single-field stars with no additional objects discernible within the resolution limit of the Spitzer telescope (\textasciitilde2 arcseconds), the potential for oversight remains. For instance, the selected objects may be binary systems or may harbour background stars within the imposed resolution limit. Furthermore, there is a possibility that the chosen objects may include young, low-mass objects that are inherently luminous in the near-infrared regions of the spectrum. Even in such cases, rigorous measures are taken to ensure the fidelity of our spectral energy distribution (SED) templates in representing standard infrared spectra corresponding to the designated spectral types.

\end{itemize}

In this study, we assume that interstellar extinction for all template objects is negligible. This assumption is justified by the fact that all objects listed in Table 2 are located within 100 pc of the Sun and therefore reside within the solar neighborhood. At such close distances, interstellar extinction is expected to be minimal. For instance, the most distant object in Table 2, SSSPM J1013–1356, has a parallax of approximately 17.7 mas, corresponding to a distance of 56.6 pc. According to the 3D dust mapping derived from Pan-STARRS 1, 2MASS, and Gaia data (\citealt{green2019}), the reddening at this distance is E(g-r)(mag) = $0.00^{+0.1}_{-0.0}$
mag, indicating negligible extinction. Given this, it is reasonable to assume that the other objects in Table 2 are similarly unaffected by interstellar extinction.

In the context of this specific investigation, we have constructed standard Spectral Energy Distribution (SED) templates encompassing spectral types ranging from M9 to L4. Accompanying this discussion are graphical representations illustrating these templates. Please refer to Figure 2 for a visual display. Within this figure, there are six plots, each showing the SED template of a different spectral type. The blue curve within the SED templates is derived from IR spectra sourced from either SpeX or NIRSPEC, while the orange curve is formulated based on IRS data. Note the vertical axis labelled 'absolute flux (W/m²),' which indicates the brightness of the object located 10 parsecs from our solar system.

\section{Analysis and Discussion}
Following the construction of Spectral Energy Distribution (SED) templates encompassing the spectral types explained in the preceding section, we compared these templates with existing photometric data to assess the validity of our models. Operating under the assumption that the most plausible spectral type for the object of interest aligns with M9.0$\pm$0.5 (\citealt{reid2008}), as discussed in Section 2, we initially employed the M9 SED templates for the comparison. The results of this investigation are depicted in Figure 3, which illustrates a plot of the template against the photometric data of the object of interest.

Photometric measurements for the object in question were sourced from the 2MASS and ALLWISE catalogs, both accessible through the online VizieR service. Furthermore, the absolute luminosity of the object was computed by leveraging parallax information obtained from the GAIA eDR3 catalog.

As shown in Figure 3, discernible disparities in absolute brightness become apparent within the J, H, K, and W3 bands, where the luminosity of the object significantly surpasses that of the template in the corresponding wavelength regions. Remarkably, the luminosity in each of these bands is approximately twice as high as that predicted by the template. Following this observation, two plausible scenarios emerge as potential explanations for this anomaly.

The initial hypothesis proposes the presence of an unresolved low-mass companion of a later spectral type. To evaluate this, we analysed archival data from the Two Micron All-Sky Survey (2MASS) and the Wide-field Infrared Survey Explorer (WISE), accessed via the Infrared Science Archive (IRSA) finder chart. Upon inspecting the images, we found no evidence of additional sources within the full-width half-maximum (FWHM) of the point spread function (PSF) across the 2MASS J, H, and K bandpasses, with FWHM values of 2.9", 2.8", and 2.9", respectively (see \citealt{IPAC2023}), and the WISE1, WISE2, and WISE3 bandpasses, with FWHM values of 6.08", 6.84", and 7.36", respectively (see \citealt{wright2010}), as shown in Figure 4. However, despite the absence of resolved companions, the observed photometric brightness of the primary object—significantly exceeding predictions from spectral energy distribution (SED) models—suggests the potential presence of an unresolved low-mass companion, which could be contributing to spectral blending with the primary source.

Evidence supporting this hypothesis comes from the GAIA eDR3 catalogue, accessed through the Vizier online tool, which reveals that the parallax of the proposed low-mass companion, Gaia DR3 6450595924275742848, is 44.9$\pm$0.6 mas, while the parallax of SIPS J2045-6332 is 44.2$\pm$0.2 mas, as shown in Table 1. The proper motion of the secondary object in right ascension and declination is approximately 78.7$\pm$0.1 mas/yr and -216.0$\pm$0.2 mas/yr, respectively, as also reported in Table 1. In comparison, the proper motion of SIPS J2045-6332 in right ascension and declination is approximately 78.1$\pm$0.4 mas/yr and -223$\pm$1 mas/yr. The angular separation between the two objects is approximately 0.3 arcseconds (projected distance $\sim$ 6.7 AU), indicating a potential gravitational connection between them. Given the uncertainties in these measurements, the similarity in parallax and proper motion values strongly suggests a high probability of gravitational binding. These findings are consistent with the conclusions of \citet{deacon2017} and \citet{marocco2013}, which is notable given that these studies predate the release of GAIA DR2. Furthermore, the computed GAIA absolute G-band magnitude ((17.89) as reported in Table 4 of \citet{kiman2019} suggests that the spectral type of the low-mass companion is likely between L3 and L4. Using the piecewise function derived by \cite{kirkpatrick2024} and the values provided in Table 16 of the same study, the mass of an L0-type dwarf is approximately 0.075 $M_\odot$. Given that the spectral type of the companion is later than L0, this value likely represents the upper limit for the low-mass companion. This estimate is in agreement with the findings of \cite{Mamajek2023}, who reported a similar mass for L2-type dwarfs, also around 0.075 $M_\odot$. No mass estimates are available for spectral types beyond L2.

To assess whether an L3 or L4 low-mass companion could account for the excess brightness previously noted, we added spectral templates of L3 and L4 to that of an M9 template, as illustrated in Figure 5. From these models, it can be seen that a low-mass companion of spectral type L3 or L4 does contribute to the observed excess brightness. However, the infrared spectra of the system alone cannot fully explain the observed excess, as the lower limits of the flux in the J, H, and K bands remain higher than the combined spectra of an M9+L3 and M9+L4 system.

To account for this discrepancy, we propose an additional hypothesis: that the system is in an early evolutionary phase. It is well-documented that young brown dwarfs typically exhibit larger radii compared to their older counterparts (e.g., \citealt{burrows1997, burrows2001}, \citealt{sorahara2013}). Therefore, if the system is indeed young, it is likely that the host star has an inflated radius, leading to an increase in brightness across all wavelengths. To investigate this hypothesis, we employed two different approaches:

1. \textbf{Identification of Young Moving Group Membership:} To determine the young moving group to which this system may belong, we employed the Banyan $\Sigma$ tool \citep{gange2018}. This online platform facilitated the input of the system's physical parameters, including right ascension (RA), declination (DEC), proper motions in both axes, radial velocity, and parallax. The radial velocity of the system, sourced from \cite{burgasser2015}, is measured at 5.0 $\pm$ 2.0 km/s. Analysis using Banyan $\Sigma$ indicated a 98.3\% probability that the system is a field star, with only a 1.5\% likelihood of membership in the Carina moving group. However, preliminary results from the forthcoming version of the Banyan $\Sigma$ tool (in preparation) suggest a 57\% probability of the system's association with the Carina moving group, and an additional 40\% probability of connection with the Columba moving group (Gagné, personal communication, 2024). Although these probabilities are not statistically definitive, there is a substantial possibility that both groups are components of a common stellar structure \citep{gange2021}, estimated to have an age of 26 $\pm$ 1.7 million years \citep{kerr2022}. Consequently, the age of SIPS J2045$-$6332 is likely to fall within this range.

2. \textbf{Analyzing surface gravity indicator}
The infrared spectra of young late M-type to early L-type dwarfs exhibit distinct physical features that are significantly influenced by gravitational effects. As these young objects undergo contraction during their early evolutionary phases, their atmospheric envelopes also contract, resulting in lower surface gravity. Consequently, the photospheric pressure decreases, leading to the emergence of these distinct features in their infrared spectra. For instance, weaker and narrower potassium (K I) lines at 1.17 and 1.25 µm, sodium (Na I) lines at 1.17 and 1.25 µm, and iron hydride (FeH) features at 0.99, 1.20, and 1.55 µm are observed. Additionally, stronger vanadium oxide (VO) absorption at 1.06 µm is evident in the infrared spectra of these young objects compared to older field objects (e.g., \citealt{lucas2001}, \citealt{allers2007}, \citealt{mcgovern2004}, \citealt{lodieu2008}, \citealt{cushing2005}). Furthermore, low-gravity objects exhibit a triangular H-band profile in the 1.5 µm to 1.6 µm range when compared to field dwarfs (e.g., \citealt{bowler2012}). This phenomenon is attributed to H\textsubscript{2} collision-induced absorption and enhanced FeH absorption (\citealt{borysow1997}, \citealt{linsky1969}, \citealt{gorlova2003}). Using gravity-sensitive indices established by \cite{allers2013}, the gravity classification of the object was quantitatively determined.

The infrared (IR) spectra of the object, observed using the FIRE instrument on the Magellan 6.5m telescope (see Figure 1), were analysed to quantitatively classify its surface gravity by computing gravity indices. This analysis employed the \texttt{classify.Gravity} function from the SPLAT Python library, following the methodology outlined by \citet{allers2013}. First, the SPLAT library was imported, and the spectra in FITS file format were loaded. The \texttt{classify.Gravity} function was then applied to compute the relevant gravity indices. The results are presented in Table 3.

\setcounter{table}{2}

\begin{table}
    \centering
    \begin{tabular}{|c|c|c|}
        \hline
        \textbf{Parameter} & \textbf{Value} & \textbf{Gravity Index} \\
        \hline
        SpT & L1.0 & - \\
        \hline
        FeH-z & $1.152 \pm 0.004$ & 1.0 \\
        \hline
        VO-z & $1.206 \pm 0.003$ & 1.0 \\
        \hline
        KI-J & $1.088 \pm 0.002$ & 1.0 \\
        \hline
        H-cont & $0.952 \pm 0.002$ & 1.0 \\
        \hline
        Gravity Class & INT-G & - \\
        \hline
    \end{tabular}
    \caption{Classification parameters and gravity indices.}
    \label{tab:gravity_indices}
\end{table}

Based on our computational analysis, we conclude that the object most likely belongs to the intermediate gravity class. If classified as such, its age is expected to range from 50 to 200 million years (\citealt{allers2013}).

Taking into account both methods for age estimation, we conclude that the age of this object is constrained to a range of 26 to 200 million years.

The analysis of surface gravity confirms that SIPS J2045$-$6332 is classified as spectral type L1, as detailed in Table 3, rather than M9, as previously suggested. Several factors may account for this discrepancy. One significant source is the inherent uncertainty associated with each measurement. For instance, the spectral type M9 determined by \citet{reid2008} using optical spectra exhibits an uncertainty of 0.5, while the measurement reported by \citet{allers2013} demonstrates an uncertainty of 1.0. Nevertheless, even when these uncertainties are combined, they remain insufficient to account for the two spectral-type difference observed between the two measurements.

Another potential contributing factor is the methodology employed by \citet{allers2013}, who used H$_2$O indices to ascertain the spectral type. As illustrated in Figure 6 of this study, there is significant scatter in the H$_2$O-1 and H$_2$O-2 index plots between M9 and L1, with considerable overlap in the index values corresponding to these spectral types. Furthermore, as previously discussed, the presence of a potential low-mass companion, with a spectral type ranging from L3 to L4, could also contribute to the observed spectral-type discrepancy. As the surface temperature decreases, the H$_2$O absorption bands in later-type objects deepen, which may influence the spectral-type measurement.

The observed spectral-type discrepancy does not affect our conclusion regarding the youth of the system. As detailed in Table 9 of the same study, our computed gravity indices classify the system within the intermediate gravity class, even if the object's spectral type is determined to be M9. Furthermore, the presence of a low-mass companion is unlikely to influence the gravity classification, as the computed gravity indices remain firmly within the intermediate gravity class, even for spectral types L3 or L4, as indicated in the same table.

\subsection{Could this system be a hierarchical triple system?}
Another plausible explanation for the overluminosity of SIPS J2045$-$6332 is its potential classification as a hierarchical triple system. The Gaia DR3 Renormalized Unit Weight Error (RUWE) for this object is 1.8 \citep{gaia2021}, which exceeds the threshold of 1.4 commonly associated with blended binary systems (\citealt{esa2018}). This suggests that SIPS J2045$-$6332 may consist of an unresolved inner binary with nearly equal mass components, accompanied by a later-type tertiary companion (Gaia DR3 6450595924275742848). The near-equal mass configuration of the inner binary would naturally result in a combined spectral energy distribution (SED) that is approximately twice as luminous as single-star templates across the full wavelength range, thereby accounting for the observed overluminosity.

\section{Conclusions}

In this study, we examined the pronounced near-infrared (IR) excess exhibited by the low-mass M9 object SIPS J2045$-$6332, particularly in the 2MASS J, H, K, and WISE 3 bands. To investigate the underlying cause of this excess, we constructed a Spectral Energy Distribution (SED) model and compared it with the photometric data of the object.

Following a thorough analysis, we concluded that the predominant factor contributing to the observed near-IR excess is the youth of the system, estimated to be between 26 and 200 million years old. Furthermore, we propose a high likelihood of a low-mass companion of mid L-type, located at an approximate projected distance of 6.7 AU from the primary star, SIPS J2045$-$6332. To definitively confirm this hypothesis, we recommend follow-up observations using large telescopes capable of providing high-resolution imagery.

\section*{Acknowledgements}
This publication makes use of data products from the Wide-field Infrared Survey Explorer, which is a joint project of the University of California, Los Angeles, and the Jet Propulsion Laboratory/California Institute of Technology, funded by the National Aeronautics and Space Administration. This research has made use of the NASA/IPAC Infrared Science Archive, which is funded by the National Aeronautics and Space Administration and operated by the California Institute of Technology.

We also utilised the 2MASS archive, a joint project between the University of Massachusetts and the Infrared Processing and Analysis Centre (IPAC) at Caltech, funded by NASA and the NSF.

This research benefits from the SIMBAD database, operated by the Centre de Données astronomiques de Strasbourg (CDS), France. We also utilised the VizieR catalogue access tool, provided by CDS, Strasbourg, France.

The national facility capability for SkyMapper has been funded through ARC LIEF grant LE130100104 from the Australian Research Council, awarded to the University of Sydney, the ANU, Swinburne University of Technology, the University of Queensland, the University of Western Australia, the University of Melbourne, Curtin University of Technology, Monash University and the Australian Astronomical Observatory. SkyMapper is owned and operated by The ANU’s Research School of Astronomy and Astrophysics. The survey data were processed and provided by the SkyMapper Team at ANU.
The SkyMapper node of the All-Sky Virtual Observatory (ASVO) is hosted at the National Computational Infrastructure (NCI). Development and support of the SkyMapper node of the ASVO has been funded in part by Astronomy Australia Limited (AAL) and the Australian Government through the Commonwealth’s Education Investment Fund (EIF) and National Collaborative Research Infrastructure Strategy (NCRIS), particularly the National eResearch Collaboration Tools and Resources (NeCTAR) and the Australian National Data Service Projects (ANDS). The Pan-STARRS1 Surveys (PS1) and the PS1 public science archive have been made possible through contributions by the Institute for Astronomy, the University of Hawaii, the PanSTARRS Project Office, the Max Planck Society and its participating institutes, the Max Planck Institute for Astronomy, Heidelberg and the Max Planck Institute for Extraterrestrial Physics, Garching, The Johns Hopkins University, Durham University, the University of Edinburgh, the Queen’s University Belfast, the Harvard-Smithsonian Center for Astrophysics, the Las Cumbres Observatory Global Telescope Network Incorporated, the National Central University of Taiwan, the Space Telescope Science Institute, the National Aeronautics and Space Administration under grant No. NNX08AR22G issued through the Planetary Science Division of the NASA Science Mission Directorate, the National Science Foundation grant No. AST1238877, the University of Maryland, Eötvös Loránd University (ELTE), the Los Alamos National Laboratory, and the Gordon and Betty Moore Foundation.

This study was based on observations taken with the IRTF/SpeX 0.8–5.5 µm Medium-Resolution Spectrograph and Imager, funded by the National Science Foundation and NASA and operated by the NASA Infrared Telescope Facility. The W.M. Keck Observatory, operated as a scientific partnership among the California Institute of Technology, the University of California, and NASA. The W.M. Keck Observatory was made possible by the generous financial support of the W. M. Keck Foundation.

The AI tool, ChatGPT, was utilised in this study exclusively for the purpose of editing and proofreading all sentences within the manuscript. It was not employed for data analysis or for the construction of spectral energy distribution (SED) models discussed herein. All sentences presented in this manuscript were composed by the authors in their own words.

{\it Software}: SPLAT (\citealt{burgasser2017}), BANYAN $\Sigma$ (\citealt{gange2018}),

{\it Facility}: The Magellan Telescopes (FIRE Echelle instrument), Spitzer(IRS)

\section*{DATA AVAILABILTY}
The Gaia data is accessible via VizieR at https://vizier.cds.unistra.fr/viz-bin/VizieR.
The spectral data from SpeX, IRS, and NIRSPEC can be obtained at the following links, respectively:

\begin{itemize}
    \item SpeX: \url{https://irtfweb.ifa.hawaii.edu/\textasciitilde spex/IRTF\_Spectral\_Library/}
    \item IRS: \url{https://irsa.ipac.caltech.edu/applications/Spitzer/SHA/}
    \item NIRSPEC:\url{https://irlab.astro.ucla.edu/BDSSarchive/}
\end{itemize}

\begin{figure*} 
\centering
\includegraphics[width=0.9\textwidth,keepaspectratio]{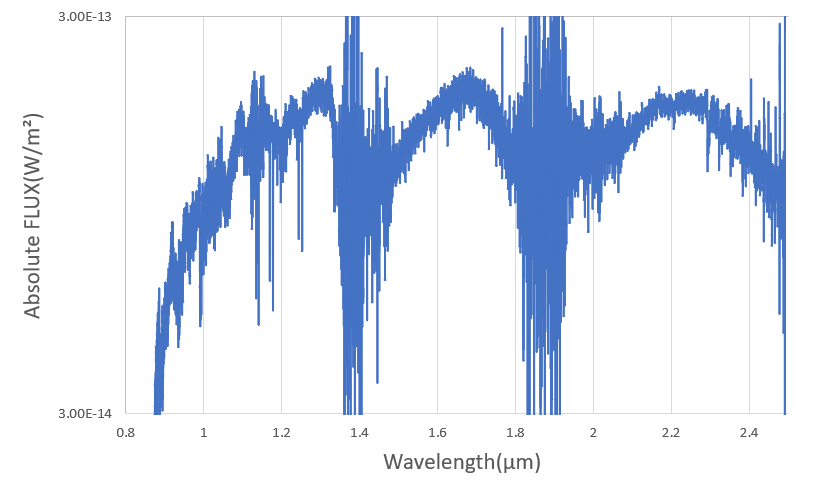}
\caption{This plot illustrates the infrared spectra of SIPS J2045-6332, as observed using the FIRE Echelle spectrograph on the Magellan telescope.
}

\label{fig:sed_comparison}
\end{figure*}

\begin{figure*}
    \centering
    
    \begin{minipage}[b]{0.48\textwidth}
        \centering
        \includegraphics[width=\textwidth, height=6cm]{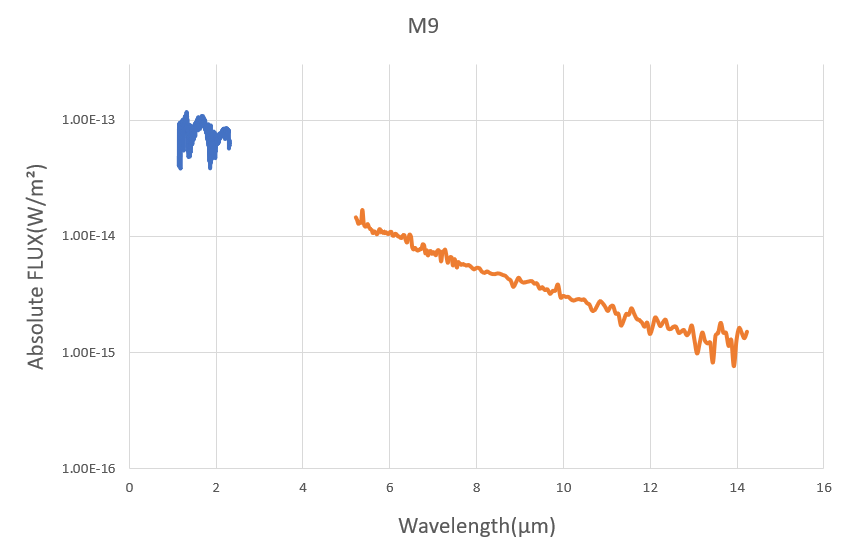}
    \end{minipage}
    \begin{minipage}[b]{0.48\textwidth}
        \centering
        \includegraphics[width=\textwidth, height=6cm]{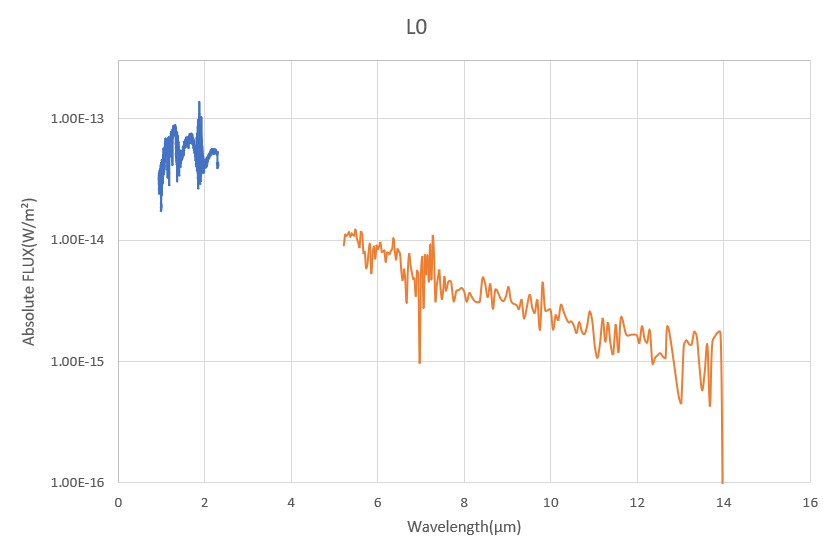}
    \end{minipage}
    
    \vspace{\baselineskip} 
    
    \begin{minipage}[b]{0.48\textwidth}
        \centering
        \includegraphics[width=\textwidth, height=6cm]{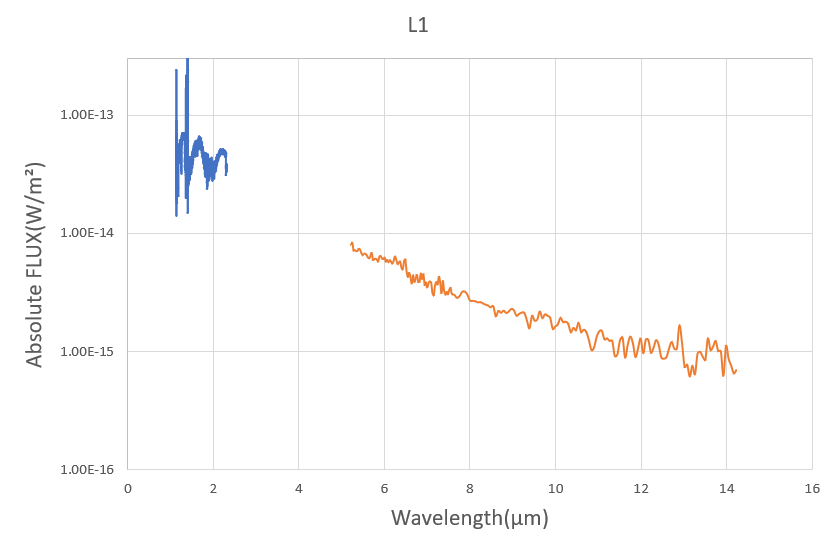}
    \end{minipage}
    \begin{minipage}[b]{0.48\textwidth}
        \centering
        \includegraphics[width=\textwidth, height=6cm]{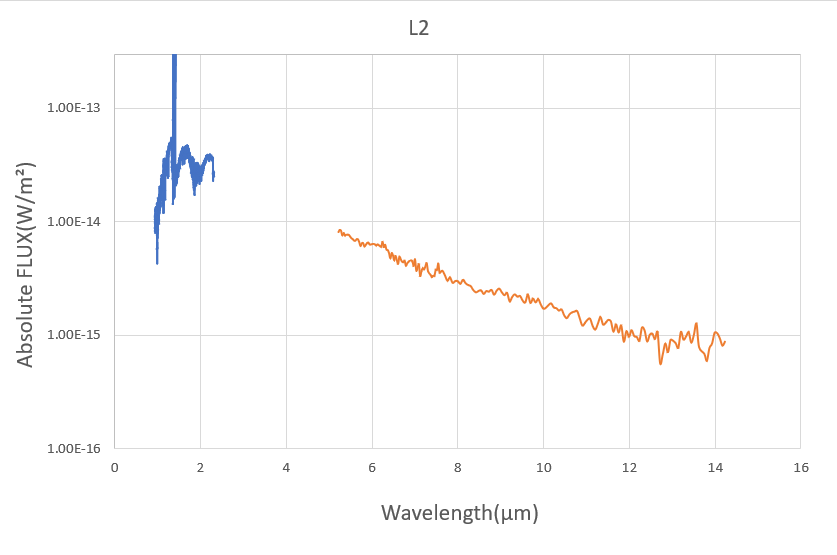}
    \end{minipage}
    
    \vspace{\baselineskip} 
    
    \begin{minipage}[b]{0.48\textwidth}
        \centering
        \includegraphics[width=\textwidth, height=6cm]{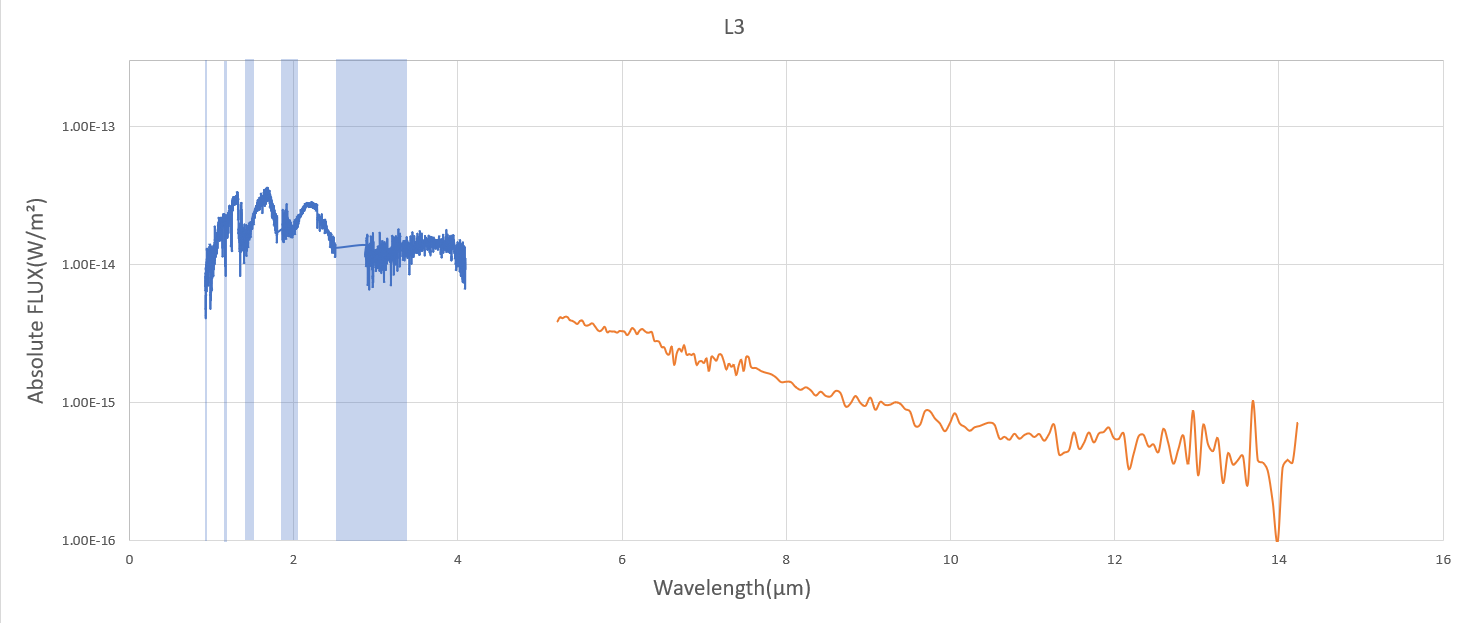}
    \end{minipage}
    \begin{minipage}[b]{0.48\textwidth}
        \centering
        \includegraphics[width=\textwidth, height=6cm]{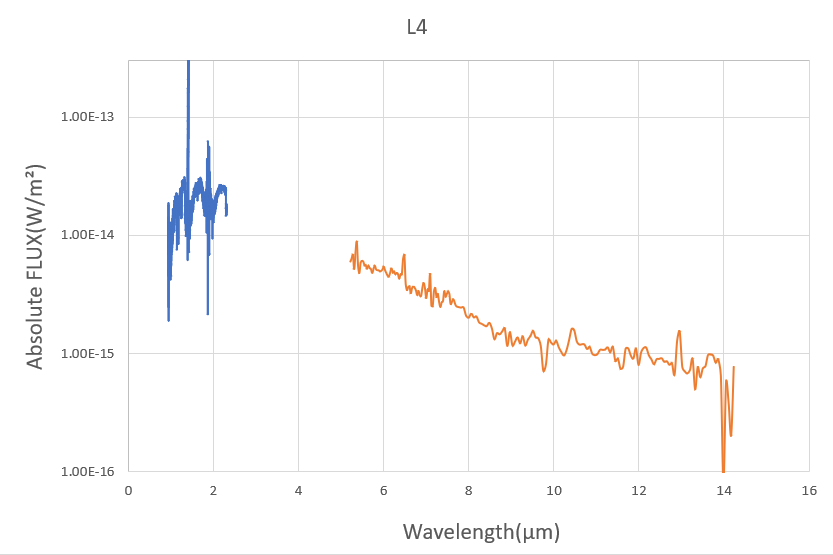}
    \end{minipage}
    
    \caption{Presented here are spectral energy distribution (SED) templates encompassing spectral types ranging from M9 to L4. The blue curve within the SED templates is derived from either IR spectra sourced from SpeX or NIRSPEC, whereas the orange curve is formulated based on IRS data. Noteworthy is the vertical axis label denoted as "Absolute Flux," indicating that the values are expressed in absolute magnitudes, thus remaining fixed.  Shaded regions in the L3 template indicate wavelengths affected by telluric absorption, where atmospheric transmission falls below 80\% (\citealt{rayner2008}).  }
\end{figure*}

\onecolumn
\setcounter{table}{1}
\begin{table*}
\centering
\caption{Physical properties of the objects used to build SED templates}\label{Table1}
\begin{longtable}{p{2.5cm} p{1.5cm} p{3cm} p{2.5cm} p{1.5cm} p{2.5cm} p{2.2cm}}
\hline
\textbf{Object ID} & \textbf{Spectral Type} & \textbf{RA and Dec} & \textbf{Plx(mas)} & \textbf{Instruments Used} & \textbf{Wavelength Covered (\textmu m)} & \textbf{Banyan Probability} \\
\hline
\raggedright LHS 2065 & M9 & 133.39836837029 
 \mbox{-03.49316442922} & 115.49$\pm$0.07 & NIRSPEC, SPEX & 1.14 - 2.31,	 0.81 - 2.41 & 99.9\% Field Star\\
\hline
\raggedright 2MASS J23515044-2537367 & M9 & 357.96210728830 \mbox{-25.62602193517} & 49.1$\pm$0.5 & IRS & 5.21-14.22 & 99.9\% Field Star \\
\hline
\raggedright 2MASS J00242463$-$0158201 & M9 & 6.10230039091 $-$01.97158639530 & 80.3$\pm$0.2 & IRS & 5.21-14.22 & 99.9\% Field Star \\
\hline
\raggedright SSSPM J1013-1356 & M9 & 153.28092302263 \mbox{-13.94381599090}	 & 17.7$\pm$0.2& IRS & 5.21-14.22 & 99.9\% Field Star\\
\hline
\raggedright 2MASS$-$J03454316+\newline 2540233 & L0 & 56.42936979352 25.67293852363 & 37.9$\pm$0.3 & NIRSPEC & 1.14 - 2.31 & 99.9\% Field Star\\
\hline
\raggedright 2MASS J09111297$+$7401081 & L0 & 137.80095069840 74.01824003360 & 40.2$\pm$0.1& IRS & 5.21-14.22 & 99.9\% Field Star\\
\hline
\raggedright 2MASSW J1035246$+$250745 & L1 & 158.85126493365 25.12772181727 & 31.4$\pm$0.4 & NIRSPEC & 1.13$-$2.31 & 99.9\% Field Star\\
\hline
\raggedright 2MASSI J1300425+191235 & L1 & 195.17379291699	19.21330233990 & 72.1$\pm$0.2 & NIRSPEC & 1.13$-$1.40 & 99.9\% Field Star\\
\hline
\raggedright 2MASS J10484281+0111580 & L1 & 162.17643932516 1.19847720791 & 66.6$\pm$0.2 & IRS & 5.21-14.22 & 99.9\% Field Star\\
\hline
\raggedright 2MASS J10511900+5613086 & L1 & 162.82732095661 56.21767229763 & 63.9$\pm$0.1& IRS & 5.21-14.22 & 99.9\% Field Star\\
\hline
\raggedright 2MASS J11083081+6830169 & L1 & 167.12534674446 68.50376157186 & 61.8$\pm$0.1 & IRS & 5.21-14.22 & 98.6\% Field Star\\
\hline
\raggedright 2MASSW J0015447+351603 & L2 & 3.93685738118 35.26625717295 & 58.7$\pm$0.3 & NIRSPEC & 0.94-2.31 & 99.9\% Field Star\\
\hline
\raggedright 2MASSI J1726000+153819 & L2 & 261.50004954533 15.63830859855 & 28$\pm$1 & NIRSPEC & 1.14-1.41 & 99.9\% Field Star\\
\hline
\raggedright 2MASS J04455387-3048204 & L2 & 71.47540368115 $-$30.80759815668 & 61.9$\pm$0.1 & IRS & 5.21-14.22 & 99.9\% Field Star\\
\hline
\raggedright 2MASS J10511900+5613086 & L2 & 162.82732095661 56.21767229763 & 63.9$\pm$0.1 & IRS & 5.21-14.22 & 99.9\% Field Star\\
\hline
\raggedright 2MASS J10452400-0149576 & L2 & 161.34766648194 $-$1.83275847502 & 58.8$\pm$0.2 & IRS & 5.21-14.22 & 99.9\% Field Star\\
\hline
\raggedright 2MASS J15065441+1321060 & L3 & 226.72148203556 13.35162520086 & 85.4$\pm$0.2& SpeX & 0.93-4.09 & 99.9\% Field Star\\
\hline
\raggedright 2MASS J16154416+3559005 & L3 & 243.93381504499 35.98086198988 & 50.2$\pm$0.3& IRS & 5.21-14.22 & 99.9\% Field Star\\
\hline
\raggedright 2MASSW J1506544+132106 & L3 & 226.72148203556 13.35162520086 & 50.2$\pm$0.2& IRS & 5.21-14.22 & 99.9\% Field Star\\
\hline
\raggedright GD 165B & L4 & 216.16216205667 9.28660191412 & 29.99$\pm$0.03 & NIRSPEC & 0.94-2.31 & 99.9\% Field Star\\
\hline
\raggedright 2MASSI J2158045-155009 & L4 & 329.51926101847 $-$15.83633595321	 & 43.1$\pm$0.9 & NIRSPEC & 1.14-1.41 & 99.9\% Field Star\\
\hline
\raggedright 2MASSI J0025036+475919 & L4 & 6.26524690 47.98866300 & 23$\pm$1 & IRS & 5.21-14.22 & 99.9\% Field Star\\
\hline
\end{longtable}

\footnotesize
\textbf{Notes.} 1. The coordinates (RA, Dec) and parallaxes (Plx) for all objects, except for 2MASSI J1726000+153819 and 2MASSI J0025036+475919, were obtained from the Gaia Early Data Release 3 \citep{gaia2021}.
2. Membership probabilities were calculated using the Banyan $\Sigma$ tool, based on the methodology described in \cite{gange2018}.
3. All near-infrared (IR) spectra obtained by NIRSPEC are drawn from the NIRSPEC Brown Dwarf Spectroscopic Survey (BDSS) \citep{mclean2003,mclean2007}; spectra obtained using SpeX are from the IRTF Spectral Library \citep{cushing2005,rayner2008}; and spectra acquired by the IRS on the Spitzer Space Telescope are sourced from \citet{houck2004}.
4. The coordinates (RA, Dec) and parallaxes (Plx) for 2MASSI J1726000+153819 and 2MASSI J0025036+475919 were obtained from \citet{liu2016} and \citet{factor2022}, respectively.
\end{table*}

\begin{figure*}\ 
\centering
\includegraphics[width=0.9\textwidth,keepaspectratio]{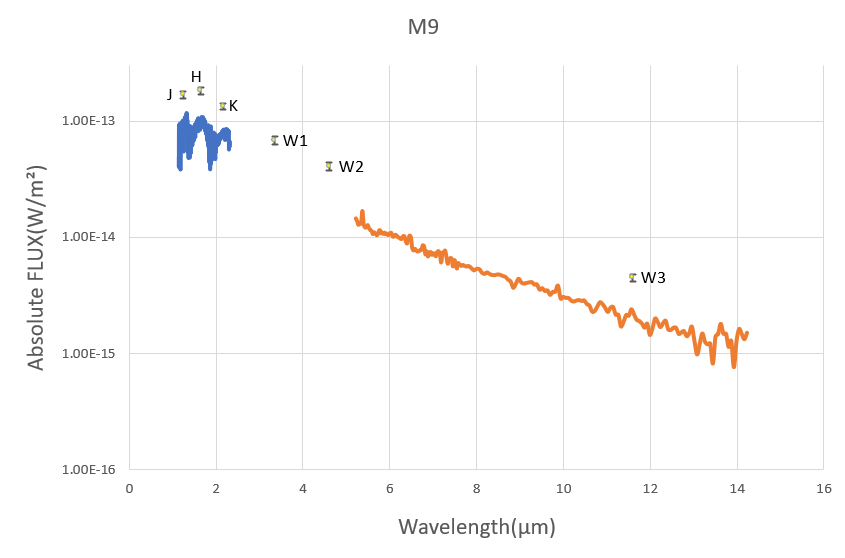}
\caption{This plot illustrates the comparison of SED template of M9 against the photometric data of the object of interest. Discernible disparities in absolute brightness become apparent within the J, H, K, and W3 bands. The luminosity in each of these bands is approximately twice as high as that predicted by the template.}
\label{fig:sed_comparison}
\end{figure*}

\begin{figure*}
\centering
\includegraphics[width=0.9\textwidth,keepaspectratio]{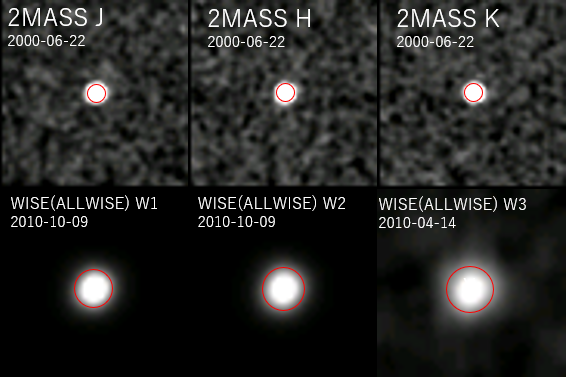}
\caption{SIPS J2045-6332 is visible in images obtained from the Two Micron All-Sky Survey (2MASS) and the Wide-field Infrared Survey Explorer (WISE), accessed via the Infrared Science Archive (IRSA) finder chart. Red circles indicate the full-width half-maximum (FWHM) of the point spread function (PSF) across the 2MASS J, H, and K bandpasses, with values of 2.9", 2.8", and 2.9", respectively (see IPAC 2023), and the WISE1, WISE2, and WISE3 bandpasses, with values of 6.08", 6.84", and 7.36", respectively (see Wright et al. 2010). These are overlaid on the image of SIPS J2045-6332. The field of view is 60x60 arcseconds. No significant source confusion or contamination is observed in any of the bandpasses; however, a strong infrared excess is detected in the 2MASS J, H, K, and WISE W3 bandpasses.}
\label{fig:second_figure}
\end{figure*}

\begin{figure*}
    \centering
    
    \begin{minipage}[b]{0.48\textwidth}
        \centering
        \includegraphics[width=\textwidth, height=6cm]{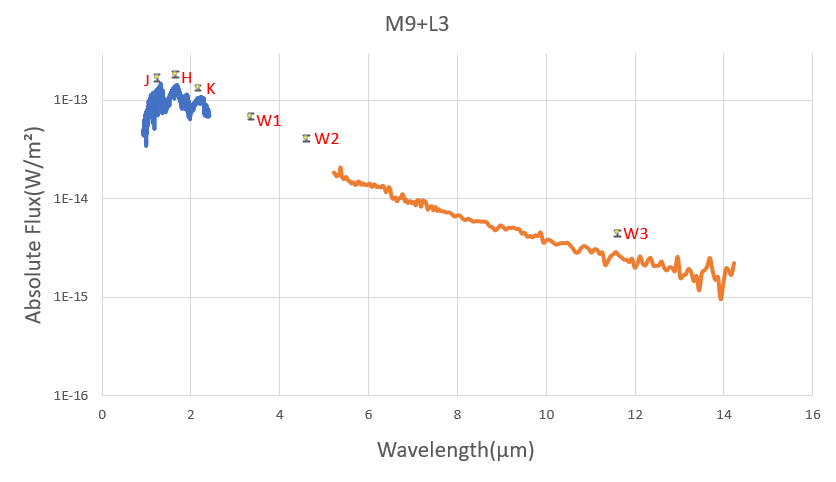}
    \end{minipage}
    \begin{minipage}[b]{0.48\textwidth}
        \centering
        \includegraphics[width=\textwidth, height=6cm]{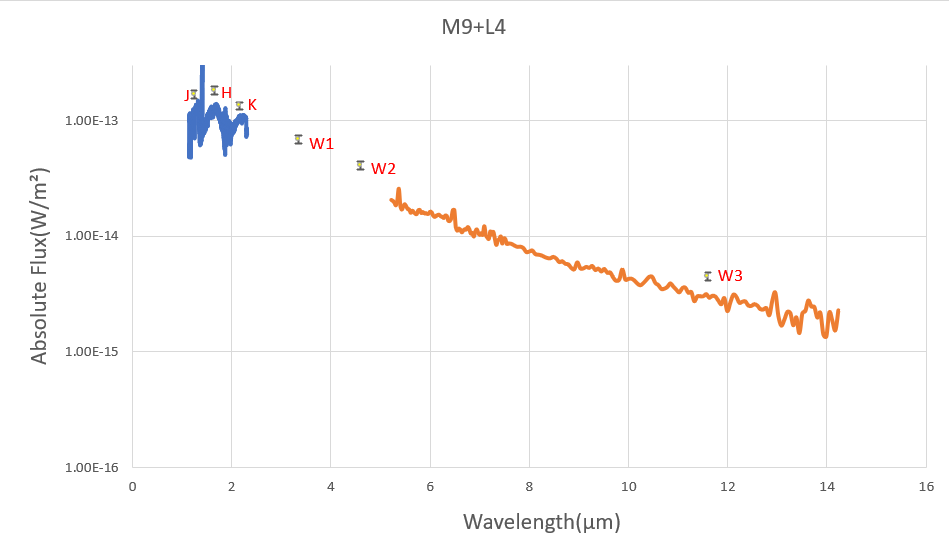}
    \end{minipage}
    
    \caption{In this figure, we present the spectral templates for M9+L3 (left) and M9+L4 (right). Despite incorporating the L3 and L4 templates into the M9 spectrum, the resulting combined brightness remains insufficient to explain the observed photometry of SIPS J2045-6332 in the J, H, K, and W3 bands. Specifically, the combined templates produce fluxes that are below the lower bounds of the observed brightness limits in all bands.}
\end{figure*}

\bsp	
\label{lastpage}
\end{document}